\documentclass[twocolumn,floatfix,aps,superscriptaddress,pra]{revtex4}
\usepackage{graphicx}
\usepackage{amsmath}
\usepackage{amssymb}
\usepackage{bm}

\begin{document}

\title{Attractor-repeller pair of topological zero-modes in a nonlinear quantum walk}
\author{Y. Gerasimenko}
\affiliation{Instituut-Lorentz, Universiteit Leiden, P.O. Box 9506, 2300 RA Leiden, The Netherlands}
\affiliation{Department of Physics, Taras Shevchenko National University of Kyiv, Volodymyrska Str. 64/13, Kyiv 01601, Ukraine}
\author{B. Tarasinski}
\affiliation{Instituut-Lorentz, Universiteit Leiden, P.O. Box 9506, 2300 RA Leiden, The Netherlands}
\author{C. W. J. Beenakker}
\affiliation{Instituut-Lorentz, Universiteit Leiden, P.O. Box 9506, 2300 RA Leiden, The Netherlands}
\date{November 2015}
\begin{abstract}
The quantum-mechanical counterpart of a classical random walk offers a rich dynamics that has recently been shown to include topologically protected bound states (zero-modes) at boundaries or domain walls. Here we show that a topological zero-mode may acquire a \textit{dynamical} role in the presence of nonlinearities. We consider a one-dimensional discrete-time quantum walk that combines zero-modes with a particle-conserving nonlinear relaxation mechanism. The presence of both particle-hole and chiral symmetry converts two zero-modes of opposite chirality into an attractor-repeller pair of the nonlinear dynamics. This makes it possible to steer the walker towards a domain wall and trap it there.
\end{abstract}
\maketitle

\section{Introduction}
A classical random walk is invariably associated with diffusive motion, but quantum superposition and interference allow for a more varied dynamics. A quantum walk can explore phase space more rapidly than its classical counterpart \cite{Aha93,Mey96,Far98}, a shift from diffusive to ballistic dynamics that is at the origin of the quadratic speed-up of quantum search algorithms \cite{Kem03,Ven12}. Diffusion is recovered for temporal disorder, while spatial disorder can induce an Anderson quantum phase transition to localized wave functions \cite{Joy10,Ahl11a,Ahl11b,Sch11,Obu11,Gho14,Edg15}. 

Two recent developments have further enriched the phenomenology: One development is the discovery that quantum walks can exhibit a topological phase transition, at which a bound state (a so-called zero-mode) appears at a boundary or domain wall \cite{Kar09,Rud09,Zah10,Kit10,Kit12,Asb14,Car15,Pol15,Zeu15}. A second development involves the introduction of nonlinearities in the dynamics \cite{Lah12,Lee15}. These have been associated with soliton structures \cite{Nav07,Mol15} and investigated as a means to speed up the quantum search \cite{Mey14}. Here we wish to connect these two separate developments, and explore how nonlinearities manifest themselves in a topological quantum walk. 

We consider the simplest case of a one-dimensional discrete-time quantum walk in the chiral orthogonal symmetry class (also known as class BDI, familiar from the Su-Schrieffer-Heeger model \cite{Su79}). The topological phase transition manifests itself by the appearance of a pair of zero-modes of opposite chirality. We demonstrate that these zero-modes may survive in the presence of nonlinearities and moreover acquire a special role as the attractor and repeller of the nonlinear dynamics.

\section{Formulation of the linear quantum walk}
We study the one-dimensional dynamics of a two-level system, represented by a spin-$\frac{1}{2}$ degree of freedom on the lattice $x\in\mathbb{Z}$. We employ a stroboscopic description, so that time $t\in\mathbb{Z}$ is discretized as well as space. The linear dynamics is obtained by repeated applications of a unitary operator $U$ on a spinor $\psi$,
\begin{equation}
\psi_t=(U)^{t}\psi_0,\;\; \psi_t(x)=\bigl(u(x,t),v(x,t)\bigr).
\end{equation}
Quite generally, a single time step of such a discrete-time quantum walk can be decomposed into two operations: A rotation $R_\vartheta$ of the spinor and a shift $S$ to the left or to the right dependent on the spin component:
\begin{align}
&R_{\vartheta}\psi=e^{-i\vartheta\sigma_{y}}\psi=(u\cos\vartheta-v\sin\vartheta,u\sin\vartheta+v\cos\vartheta),\nonumber\\
&S\bigl(u(x,t),v(x,t)\bigr)=\bigl(u(x-1,t),v(x+1,t)\bigr).
\end{align}
We can combine the two operations as $SR_\vartheta$ or $R_\vartheta S$, but we prefer to take the symmetrized product \cite{Asb13},
\begin{equation}
U=R_{\vartheta/2}SR_{\vartheta/2}.\label{Udef}
\end{equation}

The evolution operator \eqref{Udef} is representative of a chiral orthogonal quantum walk, meaning that $U=U^\ast$ is real orthogonal (particle-hole symmetry) and$(\sigma_x U)^2=1$ (chiral symmetry). This BDI symmetry class supports a topologically protected zero-mode bound to a domain wall where $\vartheta(x)$ changes sign. Its time-independent state $\Psi_\pm(x)$ satisfies \cite{note1}
\begin{equation}
U\Psi_\pm=\Psi_\pm,\;\;\sigma_x\Psi_\pm=\pm\Psi_\pm.\label{Psipmdef}
\end{equation}
The eigenvalue $\pm 1$ of the Pauli matrix $\sigma_x$ distinguishes the chirality of the zero-mode \cite{note2}.

\section{Introduction of a nonlinearity} 
We now introduce a nonlinearity (strength $\kappa$) into the quantum walk by inserting a $\psi$-dependent rotation at each time step,
\begin{subequations}
\label{UMtdef}
\begin{align}
&\psi_{t+1}(x)=U\bar{\psi}_t(x),\label{nonlinearUa}\\
&\bar{\psi}_t(x)=\exp\bigl(-i\kappa M_{z}(x,t)\sigma_y\bigr)\psi_t(x),
\label{nonlinearUb}\\
&M_z(x,t)=\psi_t^\dagger(x)\sigma_z\psi_t(x)=|u(x,t)|^2-|v(x,t)|^2.\label{Mtdef}
\end{align}
\end{subequations}
This nonlinear time-evolution conserves particle-hole symmetry (a real $\psi$ remains real), but chiral symmetry no longer applies. Still, a zero-mode $\Psi_\pm$ of the linear problem ($\kappa=0$) remains a stationary state when we switch on the nonlinearity, because $M_{z}=0$ for any eigenstate of $\sigma_x$.

To appreciate the new features introduced by the nonlinearity, it is helpful to look at a uniform $\vartheta$ and a real initial state $\psi=(\cos\alpha,\sin\alpha)$ without any spatial dependence. In one time step the angle $\alpha$ is mapped to $\alpha+\vartheta+\kappa\cos 2\alpha$. This map is invertible if $|\kappa|\leq 1/2$, but it is not area preserving. The phase space contracts around one of two attractive fixed points, defined by $\cos 2\alpha_c=-\vartheta/\kappa$, $\sin 2\alpha_c>0$. Note that this relaxation does not involve any loss of particles: $\sum_x(|u|^2+|v|^2)$ is conserved by the nonlinear dynamics.

As we will now show, for a spatially dependent $\vartheta(x)$ the zero-mode at a domain wall becomes an attractive or repulsive fixed point, depending on its chirality. We first present numerical evidence and then give the analytical solution in the continuum limit.

\section{Collapse onto a zero-mode} 
\begin{figure}[tb]
\centerline{\includegraphics[width=0.8\linewidth]{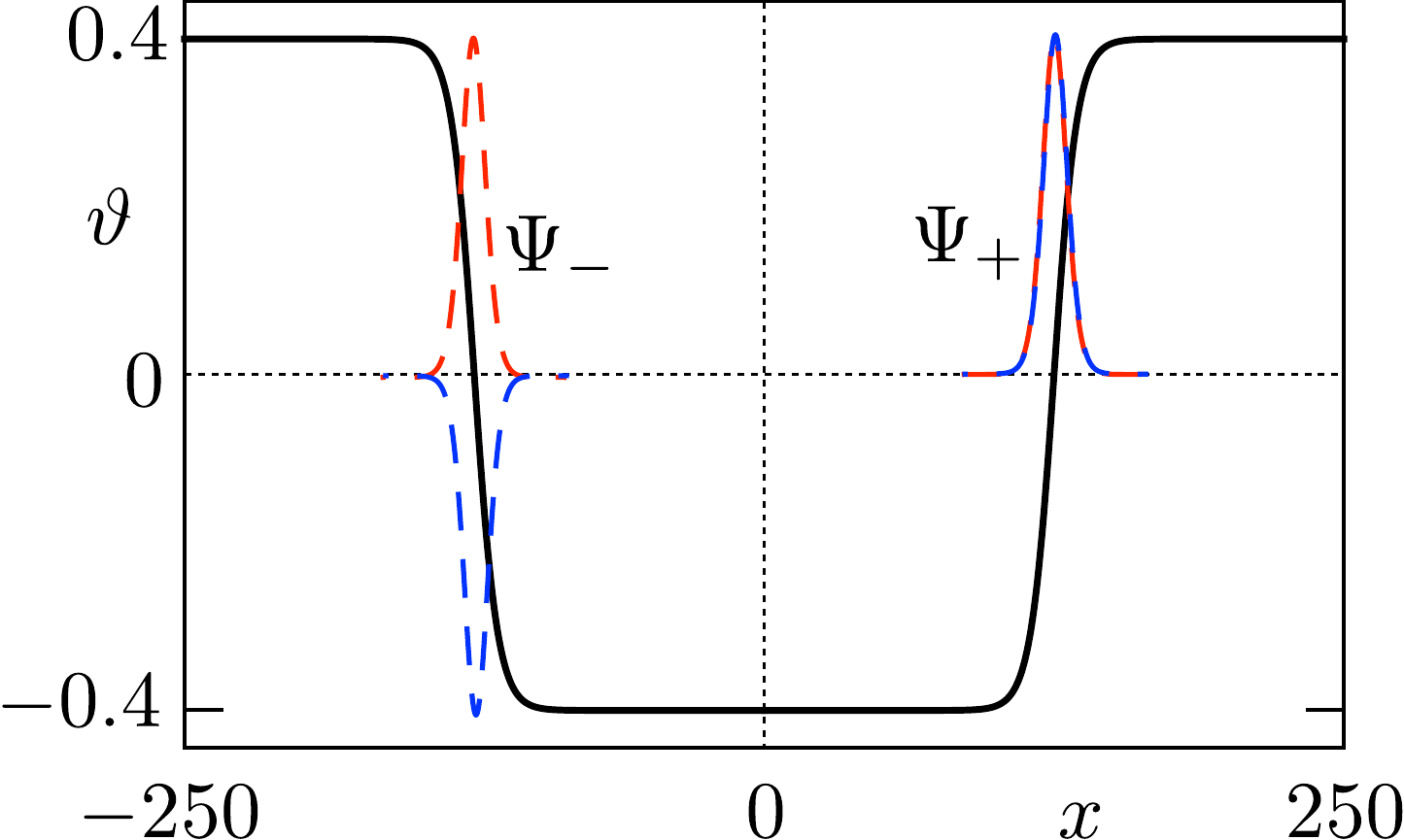}}
\caption{[color online] Solid curve: Position-dependent rotation angle $\vartheta(x)$ with a pair of domain walls at which the angle changes sign. Plotted is the profile \eqref{thetaprofile} with $L=500$, $\lambda=10$, $\vartheta_0=0.4$ used in the numerical simulations. Dashed curves: The two (unnormalized) spinor components of the zero-modes bound to the two domain walls, calculated from Eq.\ \eqref{psiuresult}. The state $\Psi_\pm$ is an eigenvector of $\sigma_x$ with eigenvalue $\pm 1$.
}
\label{fig_domainwall}
\end{figure}

\begin{figure}[tb]
\includegraphics[width=1\columnwidth]{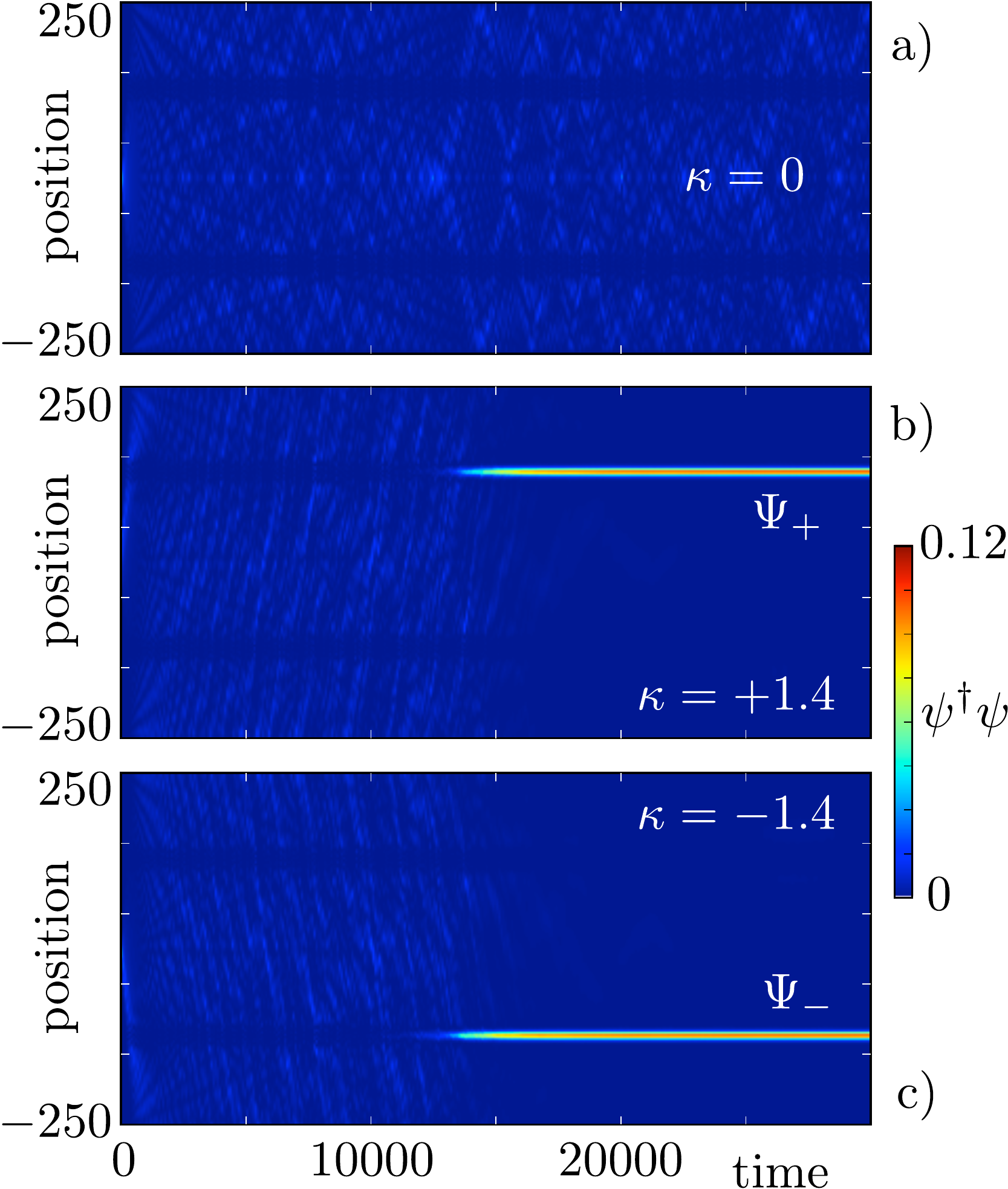}
\caption{[color online] Time-evolution of the density $\psi_t^\dagger\psi_t^{\protect\vphantom{\dagger}}$, starting from a real Gaussian wave packet $\psi_0=(u_0,u_0)$ (given by Eq.\ \eqref{psi0def} with $\sigma^2=50$), for the quantum walk with rotation angle profile of Fig.\ \ref{fig_domainwall}. The three panels show the result for the linear quantum walk (panel a, $\kappa=0$) and for the nonlinear quantum walk (panels b and c, $\kappa=\pm 1.4$). Depending on the sign of the nonlinearity, the state collapses onto the zero-mode $\Psi_+$ or $\Psi_-$.
}
\label{fig:real-init}
\end{figure}

We take a lattice of length $L$ with periodic boundary conditions, $-L/2<x<L/2$. The profile of $\vartheta(x)$ consists of two domains, with domain walls of width $\lambda\ll L$ at $x_\pm=\pm L/4$:
\begin{equation}
\vartheta\left(x\right)=\begin{cases}
\vartheta_{0}\tanh(x/\lambda-L/4\lambda) & \mbox{for}\; 0< x<L/2,\\
-\vartheta_{0}\tanh(x/\lambda+L/4\lambda) & \mbox{for}\;-L/2< x< 0,
\end{cases}\label{thetaprofile}
\end{equation}
see Fig.\ \ref{fig_domainwall}. As initial condition for the numerics we take a real Gaussian wave packet centered at $x=0$,
\begin{equation}
\psi_0=(u_0,u_0),\;\;u_0(x)=(2\sigma\sqrt\pi)^{-1/2}\exp(-x^{2}/2\sigma^{2}),\label{psi0def}
\end{equation}
normalized to unity, $\int\psi_0^\dagger\psi_0^{\vphantom\dagger}\,dx=1$. Fig.\ \ref{fig:real-init} shows how this state collapses onto one of the two domain walls, depending on the sign of $\kappa$.

For the analytics we take the continuum limit of the discrete-time quantum walk, obtained from Eq.\ \eqref{UMtdef} under the assumption that the change $\delta\psi$ in one time step $\delta t$ is infinitesimal. The state-dependent rotation contributes a term $-i\delta t(\vartheta+\kappa\psi^\dagger\sigma_z\psi)\sigma_y\psi$ to $\delta\psi$, while the state-dependent shift contributes $-\delta t\sigma_z\partial\psi/\partial x$, resulting in the Dirac equation \cite{Mey96}
\begin{equation}
i\frac{\partial\psi}{\partial t}=-i\sigma_{z}\frac{\partial\psi}{\partial x}+\bigl(\vartheta\left(x\right)+\kappa\psi^{\dagger}\sigma_{z}\psi\bigr)\sigma_{y}\psi.\label{Diraceq}
\end{equation}
For large $L$ the two domain walls may be considered separately. The zero-mode bound to the domain wall at $x_\pm=\pm L/4$ is given by
\begin{equation}
\Psi_\pm\propto(u_\pm,\pm u_\pm),\;\; u_\pm(x)=\exp\left(\pm\int^0_{x}\vartheta(x')dx'\right).\label{psiuresult}
\end{equation}
The time-independent state $\Psi_\pm$ is an eigenvector of $\sigma_x$ with eigenvalue $\pm 1$, selected by the sign of $\vartheta'(x)$ at the domain wall.

We now perform a linear stability analysis for a real perturbation $\psi(x,t)=\Psi_\pm(x)+\eta(x,t)$ of the zero-mode. To linear order in $\eta$ we have
\begin{equation}
\frac{\partial\eta}{\partial t}=-\sigma_{z}\frac{\partial\eta}{\partial x}-\vartheta(x)i\sigma_{y}\eta- 2\kappa u_\pm^{2}(x)(\pm \eta-\sigma_{x}\eta).\label{etarelax}
\end{equation}
We focus on perturbations $\eta=e^{ikx}\eta(t)$ of the zero-mode with wave number $k\gtrsim 1/\lambda$, so we may neglect the spatial dependence of $\vartheta(x)$ and $u_\pm(x)$. The resulting ordinary differential equation,
\begin{equation}
\frac{d\eta}{dt}=-\Gamma\eta,\;\;\Gamma=ik\sigma_z+i\vartheta\sigma_y+2\kappa u^2_\pm(\pm1-\sigma_x),\label{Gammadef}
\end{equation}
has relaxation matrix $\Gamma$ with eigenvalues $\mu_1,\mu_2$ given by
\begin{equation}
\begin{split}
&\mu_1=\pm 2\kappa u_\pm^2+\Delta,\;\;\mu_2=\pm 2\kappa u_\pm^2-\Delta,\\
&\Delta^2=4\kappa^2 u_\pm^4-k^2-\vartheta^2.
\label{relaxation_eigenvalues}
\end{split} 
\end{equation}
We conclude that for $\kappa>0$ the zero-mode $\Psi_+$ is an attractor (${\rm Re}\, \mu_1,\mu_2>0$) and $\Psi_-$ is a repeller (${\rm Re}\,\mu_1,\mu_2<0$), while for $\kappa<0$ the roles are interchanged.

\section{Initial states without particle-hole symmetry}
Particle-hole symmetry ensures that a real $\psi$ remains real, but we might start with an initially complex state and ask for the stability of the zero-mode under complex perturbations. Substitution into Eq.\ \eqref{Diraceq} of $\psi=\Psi_\pm+\eta+i\zeta$, with real $\Psi_\pm,\eta,\zeta$, shows that to first order in $\eta,\zeta$ the nonlinear term contains only the real perturbation:
\begin{align}
\frac{\partial}{\partial t}(\eta+i\zeta)={}&-\sigma_{z}\frac{\partial}{\partial x}(\eta+i\zeta)-\vartheta(x)i\sigma_{y}(\eta+i\zeta)\nonumber\\
&- 2\kappa u_\pm^{2}(x)(\pm \eta-\sigma_{x}\eta).
\end{align}

\begin{figure}[tb]
\includegraphics[width=1\columnwidth]{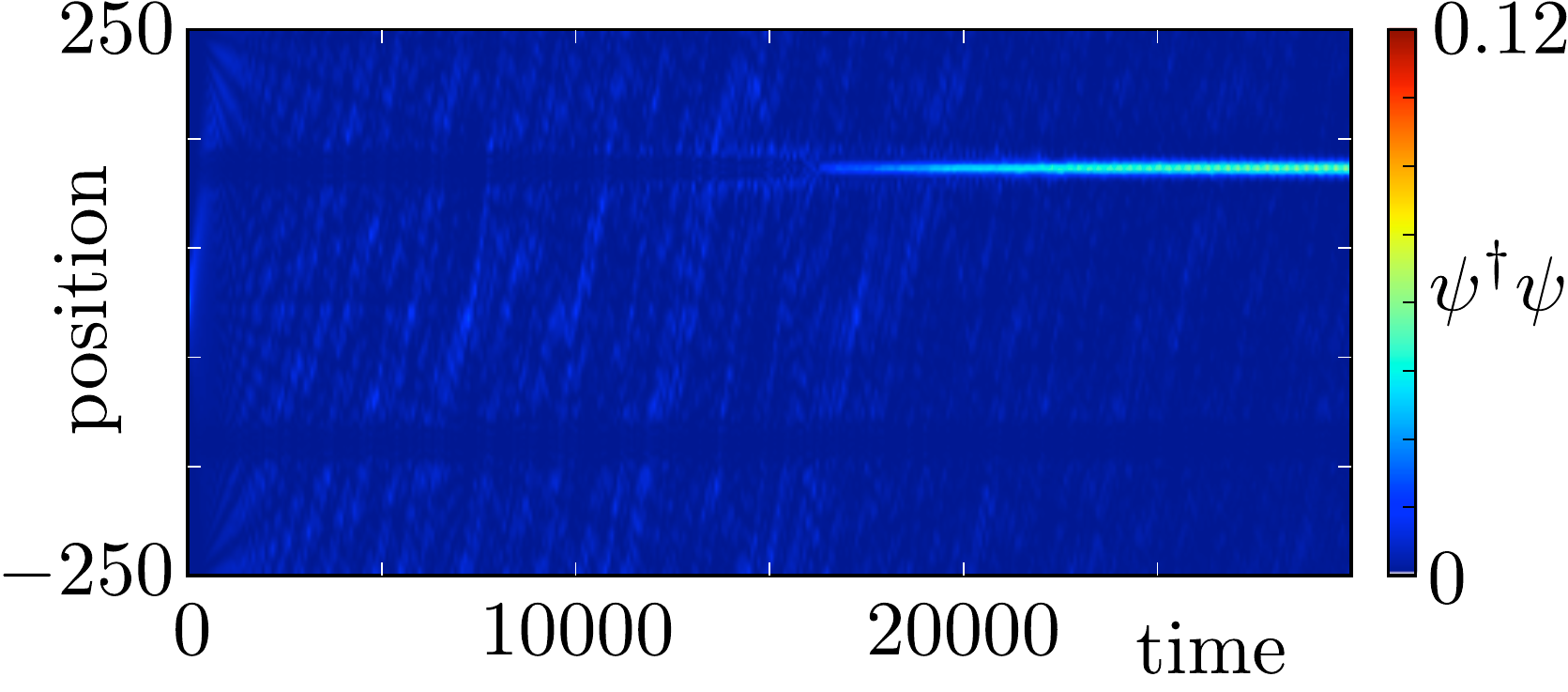}
\caption{[color online] Same as Fig.\ \ref{fig:real-init}b, but with a complex initial state $\psi_0=(u_0,iu_0)$.
}
\label{fig:complex-init}
\end{figure}

\begin{figure}[tb]
\includegraphics[width=0.9\columnwidth]{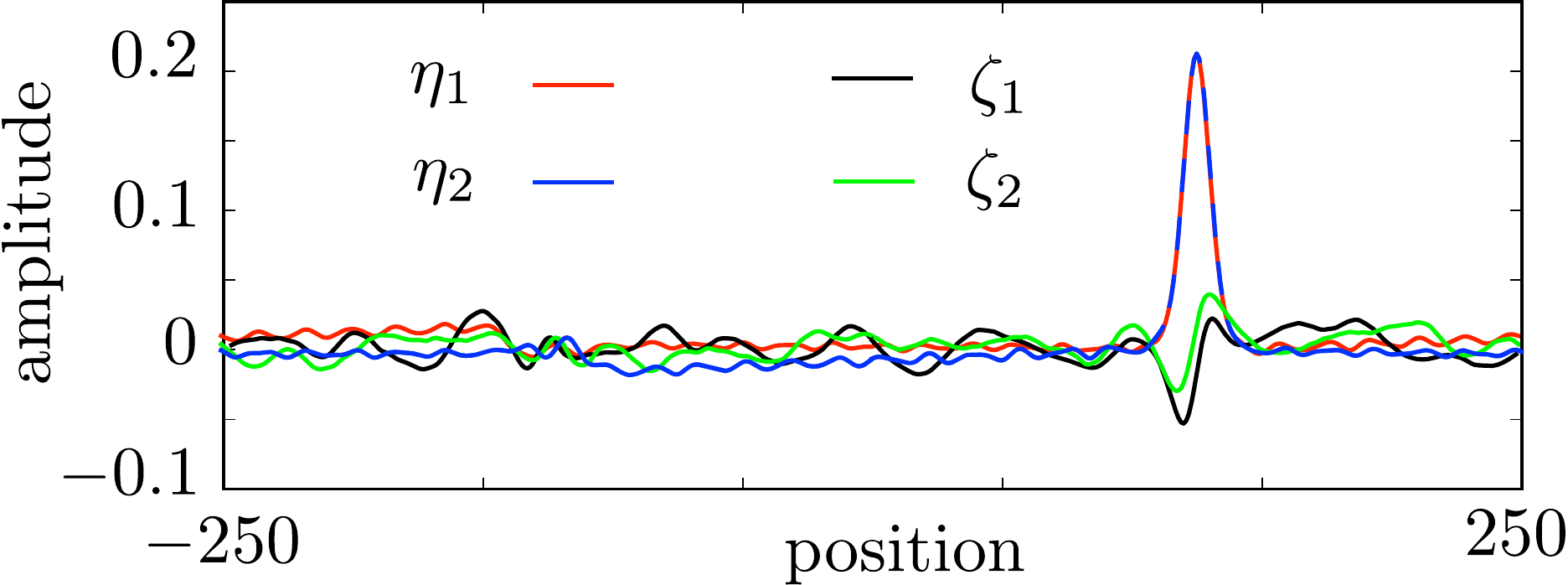}
\caption{[color online] Decomposition of the state $\psi=e^{i\phi}(\eta+i\zeta)$ at a late time ($t=8\cdot 10^4$), starting from the complex state $\psi_0=(u_0,u_0+iu_0)$, with $u_0$ the Gaussian wave packet \eqref{psi0def} ($\kappa=1.4$, other parameters as in Fig.\ \ref{fig_domainwall}). The spinor $\eta=(\eta_1,\eta_2)$ is in-phase with the zero-mode $\Psi_+$, the spinor $\zeta=(\zeta_1,\zeta_2)$ is out-of-phase.
}
\label{fig:complex-separation}
\end{figure}

The relaxation matrix for the real perturbation is as in Eq.\ \eqref{Gammadef}, with eigenvalues $\mu_1,\mu_2$ given by Eq.\ \eqref{relaxation_eigenvalues}. But the relaxation matrix for the imaginary perturbation,
\begin{equation}
\frac{d\zeta}{dt}=-\Gamma_0\zeta,\;\;\Gamma_0=ik\sigma_z+i\vartheta\sigma_y,\label{Gamma0def}
\end{equation}
has purely imaginary eigenvalues,
\begin{equation}
\mu_3=i\sqrt{k^2+\vartheta^2},\;\;\mu_4=-i\sqrt{k^2+\vartheta^2}.\label{relaxation_eigenvalues_im}
\end{equation}
More generally, a perturbation of a
complex zero-mode $\Psi_\pm(x)=e^{i\phi}(u_\pm,u_\pm)$ has (for
$\kappa>0$) a decaying in-phase component $e^{i\phi}\eta$ and a
non-decaying out-of-phase component $ie^{i\phi}\zeta$ [with real spinors
$\eta=(\eta_1,\eta_2),\zeta=(\zeta_1,\zeta_2)$]. Figs.\ \ref{fig:complex-init}
and \ref{fig:complex-separation} illustrate the resulting localized peak on
the extended background.

\section{Discussion} Fig.\ \ref{fig:real-init} summarizes our key finding:
While the linear quantum walk is only slightly perturbed by the emergence of
zero-modes at a topological phase transition, once we turn on the nonlinearity
the wave packet is steered towards a domain wall and trapped in a zero-mode
of definite chirality. This striking dynamics follows from a specific model
calculation. How generic is it, and how might it be realized in an experiment?

\begin{figure}[b]
\includegraphics[width=\columnwidth]{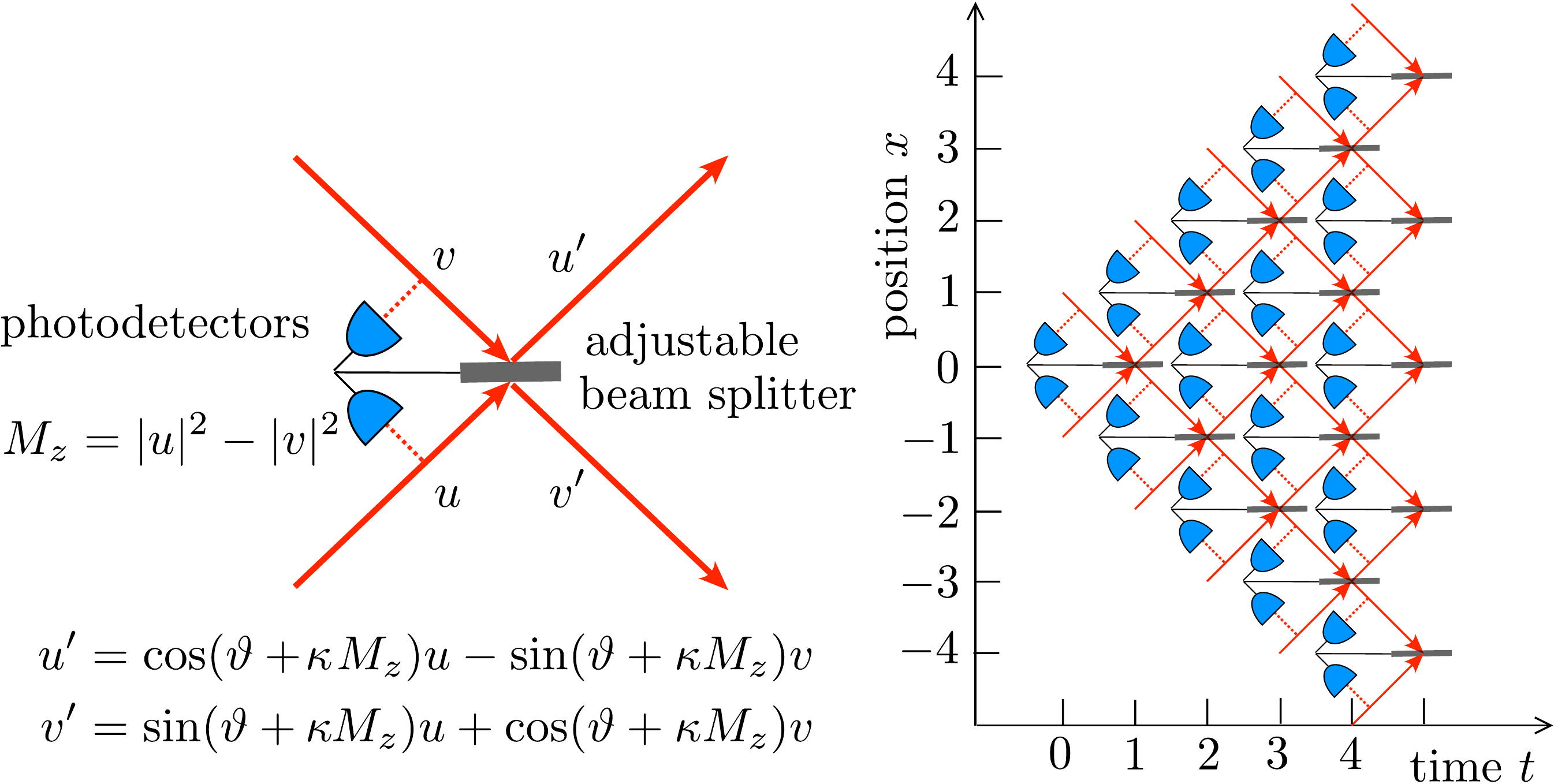} 
\caption{[color online] Optical Galton board consisting of an array of beam splitters with an adjustable transmission,
conditioned on the output of a pair of photodetectors. The left panel shows
a single element of the array, the right panel shows their combination.
}
\label{fig:Galton}
\end{figure}

For the experimental connection, we recall that quantum walks can
be realized with true quantum mechanical elements \cite{Man14} (ion
traps, cold atoms, quantum dots) --- or they can be simulated with
classical waves \cite{Kni03,Jeo03}, as in the optical Galton board
\cite{Bou00,Do05,Sch10}. Such a simulated quantum walk combines linear
optical elements to mimic the quantum evolution of a spin-$1/2$ degree of
freedom. Nonlinearities can be introduced via nonlinear optics \cite{Sol12},
or while staying within linear optics by introducing a feed-forward element
conditioned on the output of a photodetector \cite{Shi14}. A scheme of the
latter type \cite{note3} is illustrated in Fig.\ \ref{fig:Galton}. This
optical Galton board simulates a quantum walk with evolution operator
$SR_{\vartheta}\exp(-i\kappa M_z\sigma_y)$, which differs from Eqs.\
\eqref{Udef} and \eqref{UMtdef} by the order of the operators ($SR_\vartheta$
instead of $R_{\vartheta/2}SR_{\vartheta/2}$). In the continuum limit of Eq.\
\eqref{Diraceq} this order is irrelevant, and we have checked numerically
that the dynamics is essentially the same as in Fig.\ \ref{fig:real-init}.

Concerning the generality of the result, we have two necessary conditions for
the nonlinearity: it should preserve the zero-mode as a fixed point of the
dynamics and it should contract phase space, breaking the area-preservation
of the linear dynamics. Both conditions hold if Eq.\ \eqref{UMtdef} is replaced by 
\begin{equation}
\begin{split}
    \psi_{t+1} = U\bar{\psi}_t,\;\;\bar{\psi}_t= \exp (-i \tilde\kappa \tilde{M} \hat{n}\cdot\hat{\sigma} )\psi_t,\\
    \tilde{M} =\psi_t^\dagger \left( \hat{m}\cdot \hat{\sigma} \right) \psi_t,
\end{split}
\end{equation}
with $\hat{\sigma}=(\sigma_x,\sigma_y,\sigma_z)$ and two unit
vectors $\hat{n}=(0,n_y,n_z)$ and $\hat{m}=(0,m_y,m_z)$ satisfying
$\hat{m}\times\hat{n}\neq 0$ (otherwise the map would be area
preserving). Particle-hole symmetry is broken for $n_z\neq 0$, but the
zero-mode $\Psi_\pm$ is preserved. A complex perturbation $\delta\psi$ has
relaxation matrix $d\delta\psi=-\tilde{\Gamma}\delta\psi$ with eigenvalues
$\tilde\mu_{n}$, $n=1,2,3,4$, given by Eqs.\ \eqref{relaxation_eigenvalues}
and \eqref{relaxation_eigenvalues_im}, upon the replacement $\kappa\mapsto
\tilde\kappa (\hat n \times \hat m)\cdot \hat x$. The attractor-repeller
pair is preserved, demonstrating the generality of our findings.

We finally note that discrete time quantum walks 
have been used as a design principle for quantum algorithms.
For instance, the search algorithms of Refs.~\onlinecite{She03, Amb05} can be 
understood in terms of bound states in effectively one-dimensional quantum walks.
The key observations in this paper, namely the convergence towards 
certain bound states from arbitrary initial states, as well as the 
accelerated escape from unwanted bound states, thus may have promising 
implications for quantum algorithms. 
This is in line with several other recent results on continuous time quantum walks,
where non-linearities are observed to speed up quantum algorithms \cite{Mey14}.

\acknowledgments 

We acknowledge discussions on the optical implementation with W. L\"{o}ffler. This research was supported by the Foundation for Fundamental Research on Matter (FOM), the Netherlands Organization for Scientific Research (NWO/OCW), and an ERC Synergy Grant.

%

%


\begin{thebibliography}{99}
\bibitem{note1} In addition to the zero-mode with $U\Psi=\Psi$, the domain wall may also support a bound state with $U\Psi=-\Psi$. Because this state is rapidly oscillating on the scale of the lattice constant, it plays no role in the long-wave length dynamics considered here.
\bibitem{note2} The fact that the zero-mode is an eigenstate of $\sigma_x$ follows from $U\Psi=\Psi$ and $U\sigma_x\Psi=\sigma_x(\sigma_x U)^2 U^{-1}\Psi=\sigma_x\Psi$. Since the zero-mode is nondegenerate, the two states $\Psi$ and $\sigma_x\Psi$ must be linearly related.
\bibitem{note3} In the implementation of an optical Galton board shown in Fig.\ \ref{fig:Galton}, the photon polarization plays no role and the spin-$1/2$ degree of freedom of the quantum walk is fully orbital \cite{Asb14}. The adjustable beam splitter combines the rotation and shift operators $R_\vartheta$ and $S$ in a single step. Alternative split-step implementations can use adjustable polarizers for $R_\vartheta$, followed by polarizing beam splitters \cite{Do05} or birefringent displacers \cite{Kit12} for $S$.
\end{thebibliography}

\end{document}